\documentclass{article}

\usepackage{lipsum}

\newcommand\blfootnote[1]{%
  \begingroup
  \renewcommand\thefootnote{}\footnote{#1}%
  \addtocounter{footnote}{-1}%
  \endgroup
}

\title{Dimensional Affect and Expression in \\Natural and Mediated Interaction\blfootnote{Proceedings of Fechner Day vol. 23 (2007). Invited article presented at the 23rd Annual Meeting of the International Society for Psychophysics, Tokyo, Japan, 20-23 October, 2007.}}

\author{
  	Michael J. Lyons\\
  	Ritsumeikan, University\\
  	Kyoto, Japan\\
  	lyons@im.ritsumei.ac.jp
}
\date{October, 2007}
\begin{document}

\maketitle

\begin{abstract}
There is a perceived controversy as to whether the cognitive representation of affect is better modelled using a dimensional or categorical theory. This paper first suggests that these views are, in fact, compatible. The paper then discusses this theme and related issues in reference to a commonly stated application domain of research on human affect and expression: human computer interaction (HCI). The novel suggestion here is that a more realistic framing of studies of human affect in expression with reference to HCI and, particularly HCHI (Human- Computer-Human Interaction) entails some re-formulation of the approach to the basic phenomena themselves. This theme is illustrated with several examples from several recent research projects.
\end{abstract}

\section*{}
The human face is involved in various aspects of verbal and non-verbal communication: acts of speech, facial expressions, facial gestures and movements, to list several major examples. Cognitive science, psychology, neuroscience, as well as sub-fields of information processing sciences are currently aimed at the study of facially mediated communication. Some of this work is aimed at developing new techniques in computer graphics, animation, and computer vision. Probably the most widely cited motivation for this research is to provide a more natural basis for humans to interact with computer systems via personification of artificially intelligent agents. While there has been considerable research along these lines, to date relatively few human computer interface technologies have successfully employed face processing. An overview of the status of HCI applications of face processing is beyond the scope of the present article, however a review of the situation has recently been given by Bartneck and Lyons \cite{bartneck2007hci}. In the current article I will focus nearly exclusively on some aspects of facial expression processing by humans and machines and how these relate to current research in human-computer interaction.
\section*{Facial Expression Representation: \\Categories or Dimensions?}
Several controversies are associated with the most fundamental issues of facial expression research, and it has been suggested by Schiano \cite{schiano2000face}, and discussed more recently by Bartneck and Lyons \cite{bartneck2007hci}, that unresolved issues might be blocking significant progress in the development of workable HCI systems. The nature of the method used to represent facial expressions is seen as a key issue in this regard. One school of thought, famously affiliated with Ekman \cite{ekman1999basic} but dating back at least to Charles Darwin, holds that a discrete set of facial expression categories serves to communicate affective states, which, likewise, can be represented using a set of emotion categories. Another view, articulated clearly by Schlosberg \cite{schlosberg1952description}, but again having roots in older work, holds that facial expressions are better suited to representation in a space having continuous dimensions of valence (pleasure/displeasure) and arousal. These views have often been presented as being mutually exclusive.
Choice of an appropriate representation scheme is no doubt of paramount importance for the success of any facial expression system, however categorical and dimensional views are by no means incompatible in the context of their application to HCI technologies. One of my earliest studies of dimensional facial expression representation, conducted with my colleagues Miyuki Kamachi and Jiro Gyoba and reported in Lyons et al. \cite{lyons1998coding}, was the result of a larger project to build a facial expression categorization system. While studying classification methods for images of facial expressions, we explored the dimensional structure of the facial expression image data and discovered that a nonlinear two-dimensional projection of the data, captured a large proportion of the variance in our data. Furthermore, the projection dimensions reflected the well-known “circumplex” of facial expressions, itself a low- dimensional projection of empirical data from semantic differential ratings of facial expression images. The correlation between the image-processing derived and semantic-rating derived spaces was unexpectedly high and provided support for our image-filter derived representation of facial expressions, as well as for the possibly utility of a dimensional representation in classifying facial expressions. At the same time, we observed a natural clustering of facial expression images within our low-dimensional affect space into basic emotional categories of happiness, anger, surprise, and so on. This finding suggested that the concept of facial expression categories could also be a viable component of our facial expression classification system.
The findings reported in Lyons et al.  \cite{lyons1998coding} and briefly summarized above showed that both categorical and dimensional representations could be used at different stages of a facial expression classification system and guided a subsequent project to build a facial expression classification system as reported by Lyons et al.  \cite{lyons1999automatic}. The basic idea of the classification system to first process facial images with filters modeled on complex cells of primary visual cortex (area V1), then project the filter outputs into a low dimensional space learned from an ensemble of facial expression images and finally categorize expressions on the basis clusters. This system embodies dimensional and categorical approaches to facial expression representation and combines the power of both: an outcome of the project was the development of one of the early successful facial expression classifiers. The general approach of combining V1-like image filtering, dimensionality reduction and categorization has become widely used.
In addition to utility of this approach for classifying images of facial expression, the schema discussed above is helpful in thinking about how dimensional and categorical facial expression representations might relate to what happens in the brain. For example, dimensional and categorical aspects of processing may be different facets of a single neural scheme for processing facial expressions. Loosely speaking, dimensionality reduction might take places at an earlier stage of processing, to reduce the complexity, and increase the robustness of a facial expression recognition system.

\section*{Facial Expression Analysis and HCI}
\subsection*{Limitations of the current approach}
The current discussion might give the impression that facial expression classification is a solved problem as, indeed, an uncritical reader of the relevant literature in pattern recognition and computer vision research might be led to think. That impression, however, would be mistaken: the application of facial expression processing techniques to working HCI systems has so far been quite limited. This situation, which has persisted for several years, led Bartneck and Lyons \cite{bartneck2007hci} to explore possible reasons for the apparent discrepancy between the reported power of available techniques, and the lack of their use in actual HCI systems. One of the significant observations to be made is that many of the proposed applications of facial expression analysis and synthesis methods suffer from “the curse of strong AI”. In other words, research projects in this field often tacitly, but unrealistically, assume the viability of an artificially intelligent agent.
To make above remarks more concrete, consider the classic proposed goal of facial expression recognition research: a software agent, or social robot, which by analyzing images of a human face can predict that human’s affective state and behave accordingly. Simply put, the stated goal is to develop a artificial system that can read the mind by looking at the face. However, this is not the same problem as classifying images or image sequences of facial expressions. There are many differences between these two problems as defined, which relate, largely, to the role of contextual information: human social intelligence relies on an ability to understand another’s emotions and predict behaviour, however this requires the judicious integration of information from a wide range of sources. More specifically, a major limitation of facial expression recognition research to date is a consequence of the reduction of the problem to the benchmark problem of classifying facial expressions into basic emotion categories. The problem has been reduced to enable optimization of algorithms on standard databases as well as to allow meaningful comparison of various methods. Performance optimization on standard tests however does not lead to a system which is capable of reading human minds or predicting behaviour: the failure of such a project should not be surprising.
\subsection*{Reframing the problem}
While advances in pattern recognition techniques will lead to ever greater recognition rates on facial expression datasets, continuous incremental improvements are unlikely to lead to progress towards the practical use of these methods. In short, a reframing of the research problem itself is needed. I propose three suggestions, with examples drawn from my work.
\subsection*{Continuously update benchmarks}
First, if any progress is to be made towards a mind-reading machine, pattern recognition researchers should continuously update their benchmarks. Let me start by giving just one specific, concrete proposal: rather than using nominal labels in terms of basic categories, images, or image sequences should be more richly described to reflect empirical data on human perception. A suggestion given by Lyons et al. \cite{lyons1998coding} is to use semantic ratings on a set of emotion labels rather than a single emotion category. The learning tasks become more complex: multivariate regression rather than hard categorization is needed. However a continuous description based on real data will be needed to progress towards reality. Collecting such a dataset is, of course, a large project and may become one of the most significant bottlenecks to future advances.
\subsection*{A constructivist approach: Artificial Expressions}
A radical reformulation of facial expression research results from the observation that our most meaningful interactions with computers are actually human-computer-human interactions, or, in other words, machine-mediated human-interactions. Why not leave the strong-AI problem of understanding emotions to humans and use machines for tasks they can perform well – reproducing, processing, and displaying information over communication networks? Systems for supporting empathetic interaction using biosensors, simple animations, and client/server modules have been proposed by Lyons et al. \cite{lyons2004facial}. Part of the approach involved scaffolding “Artificial Expressions”: easily interpretable displays of physiological variables. Meaning is attached constructively to these new expressions, through ongoing interaction between users. An analogous approach could be taken with facial expressions.
\subsection*{Feasible applications of facial expression processing}
It is not widely enough recognized that the technology developed for facial expression recognition is readily applicable to projects which are both interesting and tractable. Several examples were given by Lyons \cite{lyons2004facial} which proposed the development of “Facial Gesture Interfaces”, systems that allow motor actions of the face to be used for direct, intentional interaction with machines. Several demonstration systems have been developed. One such system allows a user to control a cursor with slight movements of the head, while entering mouse clicks by opening and closing the mouth. Another maps movements of the mouth to control a musical synthesizer or audio effects unit. A similar system uses the mouth to control brush properties for digital painting. Robust, real-time, functioning prototypes of these systems have been developed, demonstrating current technology that has adequate power for such applications. A most important aspect of these projects is that they are not cursed with the seemingly ever receding feasibility of creating artificially intelligent agents.

\bibliographystyle{ieeetr}
\bibliography{lyons_fechner_2007}

\end{document}